\documentclass[letterpaper]{ptephy_v1}

\preprintnumber{XXXX-XXXX} 

\usepackage{arydshln}
\usepackage{color}
\usepackage{multirow}

\begin{document}

\title{Development of the measurement for radium using germanium detector with molecular recognition resin}


\author[1,*]{S. Ito}
\affil{Okayama University, Faculty of Science, Okayama 700-8530, Japan \email{s-ito@okayama-u.ac.jp}}

\author[2,3*]{K. Ichimura}
\affil{Kamioka Observatory, Institute for Cosmic Ray Research, University of Tokyo, Kamioka, Gifu 506-1205, Japan \email{ichimura@km.icrr.u-tokyo.ac.jp}}
\affil[3]{Kavli Institute for the Physics and Mathematics of the Universe (WPI), the University of Tokyo, Kashiwa, Chiba, 277-8582, Japan}

\author[4]{Y. Takaku}
\affil[4]{Institute for Environmental Sciences, Department of Radioecology, Aomori, 039-3212, Japan}

\author[2,3]{K. Abe}
\author[2,3]{M. Ikeda}
\author[2,3]{Y. Kishimoto} 


\begin{abstract}%

High-purity germanium (HPGe) detectors are widely used for the measurement of the low concentrated radioactivities. 
For the Super-Kamiokande-Gadolinium project, the concentration of radium in Gd$_2$(SO$_4$)$_3{\cdot}$8H$_2$O is determined by using HPGe detectors. 
The amount of the sample is generally limited by the size of the sample space of a HPGe detector. 
This leads the limitation of the detector efficiency. 
A new method of chemical extraction using the molecular recognition resin was developed to minimize this problem. 
Using the developed method, radium could be extracted from Gd$_2$(SO$_4$)$_3{\cdot}$8H$_2$O and concentrated to smaller amount of the resin at a recovery rate of 81$\pm$4\%, which was estimated by using  barium as a recovery monitor. 
This made it possible to set the sample closer to an HPGe crystal and reduce self screening effect, resulting in increasing the detection efficiency of an HPGe detector. 
The developed method is reported on this letter. 

\end{abstract}

\subjectindex{XXX}

\maketitle

\section{Introduction}

The Super-Kamiokande-Gadolinium (SK-Gd) project is a new experiment where 0.2\% gadolinium sulfate octahydrate (Gd$_2$(SO$_4$)$_3{\cdot}$8H$_2$O) (Ref. \cite{Gd}) will be loaded into the Super-Kamiokande water Cherenkov detector (Ref. \cite{NIMA}). 
One of the main physics targets of SK-Gd is to discover supernova relic neutrinos and study star formation of the universe (Ref. \cite{SRN}). 
Measurements of solar neutrinos with a low energy threshold of $\sim$3.5 MeV (Ref. \cite{solar}) will be continued in SK-Gd therefore the several radio-impurities, $^{226}$Ra, $^{238}$U, and $^{232}$Th, in Gd$_2$(SO$_4$)$_3{\cdot}$8H$_2$O, should be minimized before loading into SK. 
The measurement of $^{226}$Ra is focused on in this letter, and see Ref. \cite{UTh} for $^{238}$U and $^{232}$Th in more detail. 
The maximum allowed level of $^{226}$Ra in Gd$_2$(SO$_4$)$_3{\cdot}$8H$_2$O for the solar neutrino analysis should be 0.5 mBq($^{226}$Ra)/kg(Gd$_2$(SO$_4$)$_3{\cdot}$8H$_2$O) (Ref. \cite{icrc}). 
This value was evaluated by comparing the energy spectrum of the solar neutrinos with $\beta$-ray from $^{214}$Bi, which is the daughter of $^{226}$Ra, around the energy threshold at 3.5 MeV. 

For SK-Gd, high-purity germanium (HPGe) detectors are used to determine such a low concentration of $^{226}$Ra in Gd$_2$(SO$_4$)$_3{\cdot}$8H$_2$O. 
The HPGe detectors are widely used for the radioactive measurement thanks to their high energy resolution, ability to identify radioactive isotopes and easy sample preparation. 
However, the amount of the sample is generally limited by the size of the sample room of the HPGe. 
This leads the limitation of the detector efficiency of the HPGe detector. For SK-Gd, although using the HPGe detectors with large sample room, the maximum amount of Gd$_2$(SO$_4$)$_3{\cdot}$8H$_2$O is 5 kg. Therefore, in spite of using the sensitive HPGe detector with large sample space, it takes about one month to achieve the precision level of 0.5 mBq/kg of $^{226}$Ra. 
In order to solve this problem, the chemical extraction is often used to concentrate radioactivities to the smaller volume sample. 
For example, the authors in Ref. \cite{Sr01} used the molecular recognition resin named ``AnaLig-Sr01", which is the product of IBC Advanced Technologies (Ref. \cite{AnaLig}), and determined the concentration of $^{226}$Ra included in rocks or building materials with the range of 5.2 to 165.0 Bq/kg. 
Since the ion radii and chemical features are very similar among strontium (Sr), radium (Ra), and barium (Ba), the resin can adsorbs not only Sr but also Ra and Ba. 
As shown in (Ref. \cite{Sr01,RaBa}), Ba is often used to monitor the recovery rate of Ra (generally called ``recovery monitor"). 

\begin{figure}[htbp]
\centerline{\includegraphics[width=10cm]{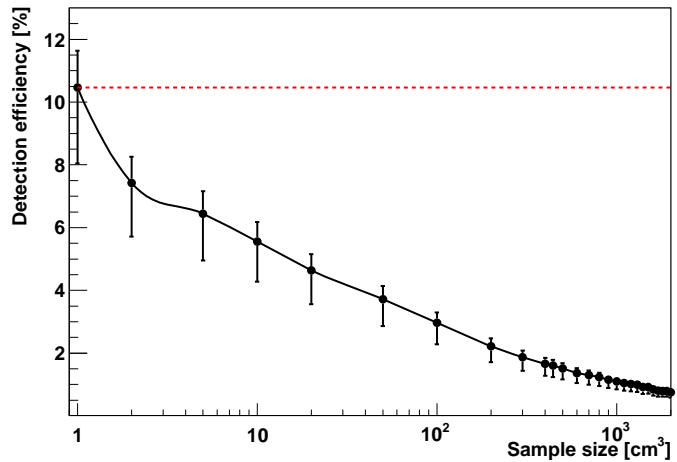}}
\caption{Result of the detection efficiency versus sample size evaluated by the Monte Carlo simulation. The horizontal axis represents sample size and the vertical axis shows the detection efficiency of Ge detector. The dashed horizontal red line indicates the concentrated sample to the resin.}
\label{fig:efficiency}
\end{figure}

The chemical extraction also has another benefits on the HPGe detector measurement. 
In general, the sensitivity of the HPGe detector highly depends on the detection efficiency related to the geometry of a Ge crystal, the sample geometry, and material due to the detector acceptance and the shielding of the samples themselves. 
Figure \ref{fig:efficiency} shows the sample size dependence of the detection efficiency for gamma ray with energy of 352 keV evaluated by the Monte Carlo simulation based on Geant4 (Ref. \cite{geant4}). 
If Ra in 500 g of Gd$_2$(SO$_4$)$_3{\cdot}$8H$_2$O, which corresponds to 440 cm$^3$, is concentrated to 2.0 g less than 1 cm$^3$) of the resin, the detection efficiency would be improved by a factor of five and shorter measurement period would be expected.

The authors in Ref. \cite{Sr01} used the column method for the chemical extraction. 
However, the amount of the sample is limited by the volume of the column. 
In our case, the maximum allowed level of $^{226}$Ra is 0.5 mBq/kg, which is four orders lower than the concentration range in Ref. \cite{Sr01}. 
In order to deal with larger amounts of the sample and enhance the concentration rate, a new chemical extraction of Ra from Gd$_2$(SO$_4$)$_3{\cdot}$8H$_2$O using batch method was developed.

\section{Chemical equipment}

The molecular recognition resin used for this study was ``AnaLig Ra-01", which is a commercially available product of IBC Advanced Technologies (Ref. \cite{AnaLig}). 
It is in the same series as ``AnaLig Sr-01", but has stronger retention for Ra. 
This resin is made of crown ethel bedded on silica support and strongly retains ions with particular ion radii. 
The resin is used for the extraction of radium (Ra) from interfering substances (Gd$_2$(SO$_4$)$_3{\cdot}$8H$_2$O in this study) and the adsorptivity of Ra with the resin does not depend on pH. 
While Ra can be eluted from the resin using ethylenediaminetetraacetic acid (EDTA) solution with pH of $\sim$11. 

In order to produce solutions with low contamination, ultra-pure SK water (Ref. \cite{NIMA}) was used for this study. 
Electronic (EL) grade of 70\% nitric acid (HNO$_3$) (Wako Pure Chemical Industries Ltd., Ref. \cite{Wako}) was also used to wash the resin and efficiently dissolve Gd$_2$(SO$_4$)$_3{\cdot}$8H$_2$O in the SK water. 
A concentration of 0.2 mol L$^{-1}$ sodium hydrogen EDTA (Na$_2$H$_2$EDTA) (Wako Pure Chemical Industries Ltd., Ref. \cite{Wako}) solution was used to wash the resin before experiment. 
Guaranteed regent grade of 25\% ammonia (NH$_3$) solution (Wako Pure Chemical Industries Ltd., Ref. \cite{Wako}) was used to adjust pH of Na$_2$H$_2$EDTA solution at $\sim$11. 

In order to easily trace the concentration of $^{226}$Ra in Gd$_2$(SO$_4$)$_3{\cdot}$8H$_2$O, $^{226}$Ra rich Gd$_2$(SO$_4$)$_3{\cdot}$8H$_2$O produced by doping hot spring water ($\sim$0.5 Bq/L) was used. 

As described in the previous section, Ba has very similar chemical feature and ionic radius with those of Ra, thus, Ba is often used as a recovery monitor of the Ra analysis. 
1000 mg L$^{-1}$ Ba of standard solution (Merck Ltd. Ref. \cite{Merck}) was used for this study. 

The inductively coupled plasma-mass spectrometry (ICP-MS) ``Agilent 7900" (Ref. \cite{Agilent}) was used to measure the concentration of Ba to estimate the recovery rate of $^{226}$Ra. 
Typical performance of this ICP-MS is described in Ref. \cite{UTh}.

\section{High-purity germanium detector}

The HPGe detector used for this analysis was a coaxial p-type HPGe crystal manufactured by CANBERRA France (Ref. \cite{canberra})
The dimension of the sample chamber was 23 $\times$ 23 $\times$ 48 cm. 
The HPGe detector was located at the center of the  sample chamber. 
 
The size of the sample space is large so that Ra with or without extraction procedure can directly be compared with each other. 
For a 500 g sample without Ra extraction, the procedure used was to put the sample in a  polytetrafluoroethylene (PTFE) bottle with 70 mm diameter and 115 mm height. 
For the resin samples with the extraction procedure, samples were enveloped by a filter paper and concentrated to a small geometry (20 mm $\times$ 20 mm $\times$ 2 mm). 
Figure \ref{fig:Ge} shows the picture of this sample measurement. 
As shown in the figure, the samples measured by the HPGe detector were covered by an ethylene-vinyl-alcohol (EVOH) bag so that Rn from samples could not penetrate. 

The concentration of $^{226}$Ra was evaluated using the characteristic $\gamma$-lines of $^{214}$Pb (609 keV) and $^{214}$Bi (352 keV and 1764 keV) with consideration of their branching ratios and detection efficiencies. 
The detection efficiency was evaluated by the Monte Carlo study. 
For example, the detection efficiency of 352 keV gamma ray originating from $^{214}$Pb, which is the daughter nuclei of $^{226}$Ra were found to be 1.7\% for 500 g of Gd$_2$(SO$_4$)$_3{\cdot}$8H$_2$O. 
On the other hand, the detection efficiency of 352 keV gamma ray for the Ra extracted and concentrated sample to 2.0 g of the resin, was evaluated to be 10.5\% due to the smaller volume of the sample. 
The procedures and results of the measurement are discussed in the following section. 

\begin{figure}[htbp]
\centerline{\includegraphics[width=9cm]{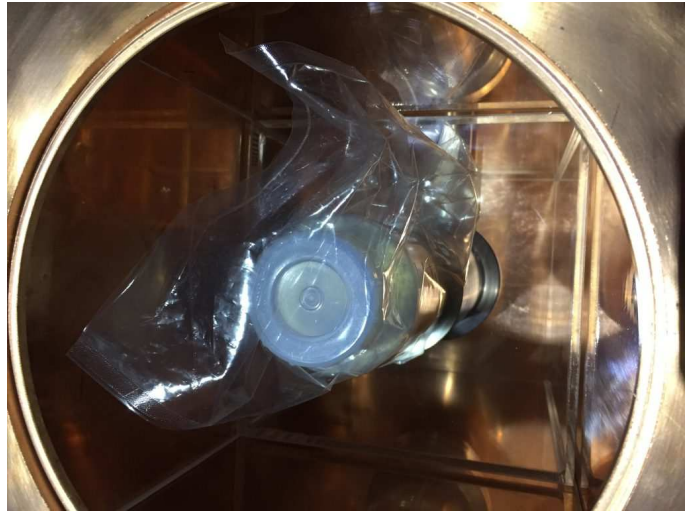}}
\caption{Picture of the HPGe measurement for 500 g of Gd$_2$(SO$_4$)$_3{\cdot}$8H$_2$O.}
\label{fig:Ge}
\end{figure}

\section{Procedure of extraction for $^{226}$Ra  from Gd$_2$(SO$_4$)$_3{\cdot}$8H$_2$O using batch method}

Initially 10 g of the resin was soaked in 50 mL of 0.1 mol L$^{-1}$ HNO$_3$ solution for one hour for prewash. 
After this procedure, the resin was washed with the ultra-pure SK water. 
Then, the resin was soaked in 50 mL of 0.05 mol L$^{-1}$ Na$_2$H$_2$EDTA solution with pH of $\sim$11 for one night, and washed with the ultra-pure SK water. 
The contamination of $^{226}$Ra in the resin was measured using the HPGe detector; the values before and after prewash were $2.56{\pm}0.34$ mBq ($^{226}$Ra)/10 g (resin) and $0.80{\pm}0.18$ mBq ($^{226}$Ra)/10 g (resin), with measurement period of 6.8 and 9.2 days, respectively.

An amount of 2.0 g of the prewashed resin was charged into the column with an inside diameter of 9 mm, and washed with 40 mL of 0.05 mol L$^{-1}$ Na$_2$H$_2$EDTA solution and 20 mL of the ultra-pure SK water. 
A beaker with a volume of 5 L, was washed using 10\% HNO$_3$ solution and the ultra-pure SK water. 
100 g of Gd$_2$(SO$_4$)$_3{\cdot}$8H$_2$O was dissolved in 900 g of 10\% HNO$_3$ solution, namely which corresponds to 1 kg of 10\% Gd$_2$(SO$_4$)$_3{\cdot}$8H$_2$O solution. 
The amount of 5.0${\times}$10$^{-5}$ g of Ba was also added into this solution as a recovery monitor (concentration of 5.0$\times$10$^{-8}$ g g$^{-1}$ in the solution). 
The washed resin and the solution were loaded into the beaker and stirred using a magnetic stirrer for one hour. 
During this stirring procedure, $^{226}$Ra and Ba were adsorbed in the resin. 

The solution including the resin was filtrated, and the resin was collected and packed into the EVOH bag, and measured using the HPGe detector. 
Figure \ref{fig:Ge} shows the energy spectra around the interested peaks. 
The concentration of remaining Ba in the solution after the batch procedure was measured using the ICP-MS. 
In order to reduce matrix effects from Gd (see Ref. \cite{UTh} in more details), the solution was diluted by a factor of 1000.

\begin{figure}[htbp]
\includegraphics[width=7.5cm]{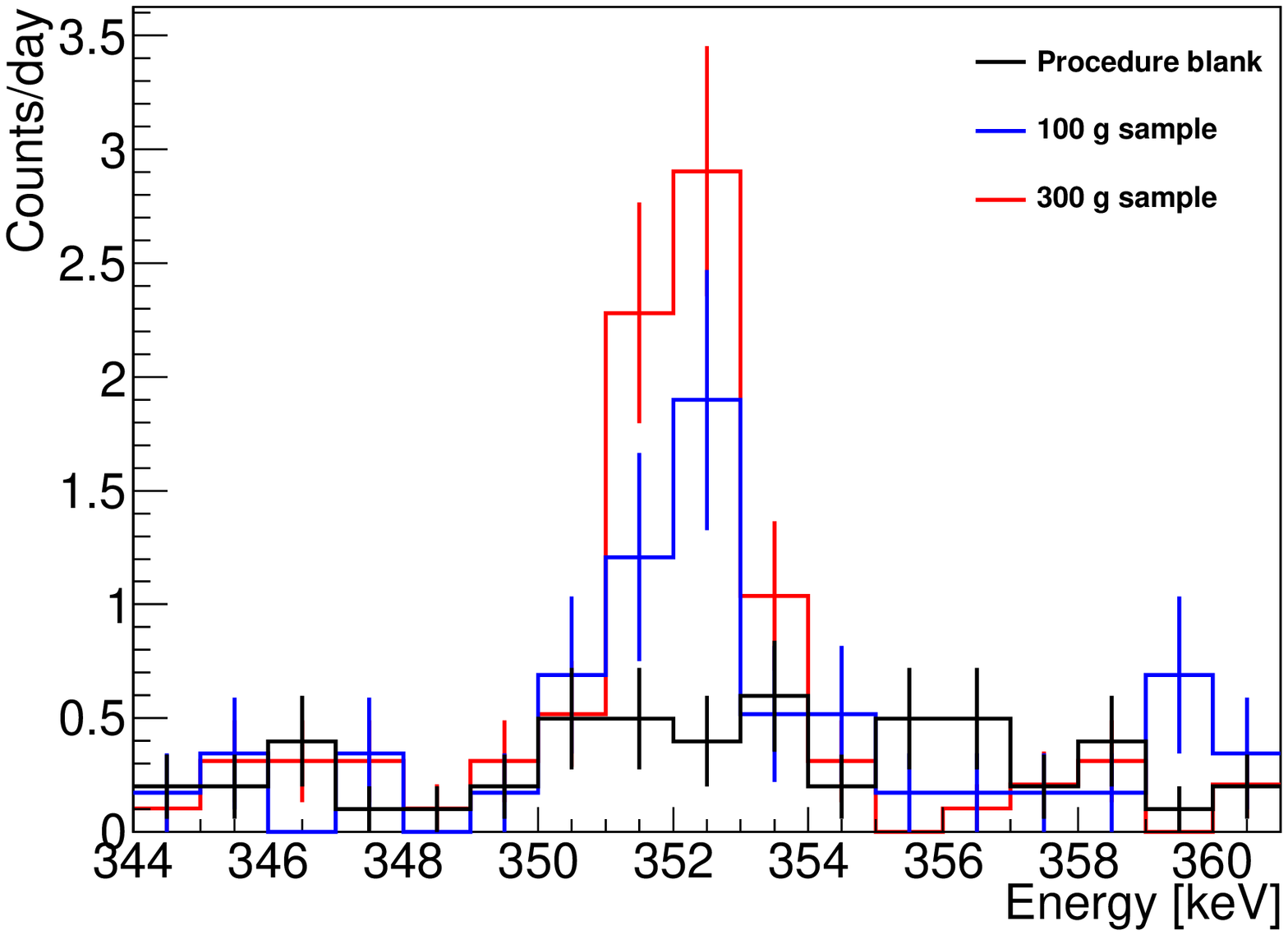}
\includegraphics[width=7.5cm]{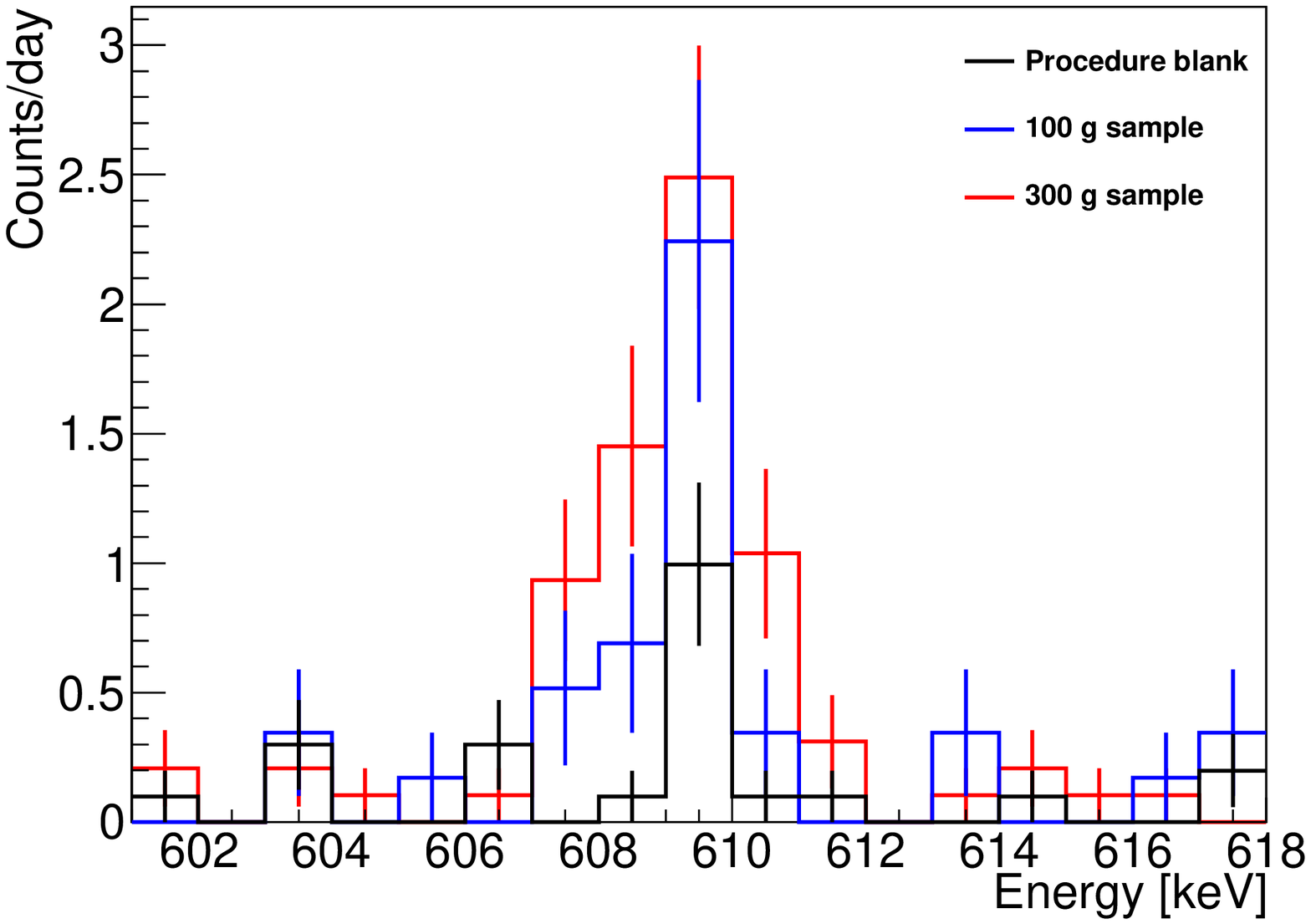}
\caption{Energy spectra of the HPGe detector around $^{214}$Bi 352 keV (left) and $^{214}$Pb 609 keV (right). Each color corresponds to 2.0 g of the resin without Gd$_2$(SO$_4$)$_3{\cdot}$8H$_2$O  (procedure blank, black) and the extracted $^{226}$Ra for 100 g (blue) and 300 g (red) samples using the resin. The error bar at each bin represents only statistical uncertainty.}
\label{fig:Ge}
\end{figure}

\section{Results of the batch study}

\begin{figure}[htbp]
\centerline{\includegraphics[width=10cm]{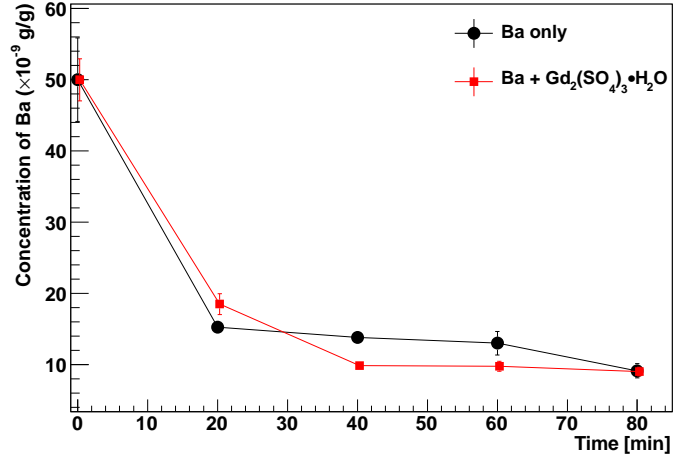}}
\caption{Result of Ba adsorptivity. The horizontal axis shows time and vertical axis represents the concentration of Ba in the solution.}
\label{fig:Ba}
\end{figure}

At first, the adsorptivity of $^{226}$Ra for the resin in the batch procedure was studied using two 5.0$\times$10$^{-8}$ g g$^{-1}$ Ba solutions, with or without Gd$_2$(SO$_4$)$_3{\cdot}$8H$_2$O. 
The solution was sampled every twenty minutes and the concentration of Ba was measured using the ICP-MS. 
Figure \ref{fig:Ba} shows the results of this study. 
For the batch study, more than twenty minutes of stirring should be performed for high adsorptivity. 
At sixty minutes of stirring, 74$\pm$9\% of Ba was recovered from the solution without Gd$_2$(SO$_4$)$_3{\cdot}$8H$_2$O. 
While 81$\pm$5\% of Ba could be extracted from 10\% Gd$_2$(SO$_4$)$_3{\cdot}$8H$_2$O solution. 
From this study, the interference from Gd$_2$(SO$_4$)$_3{\cdot}$8H$_2$O to the retention of the resin was not seen in the batch procedure. 
Stability of the recovery rate was also studied by ten times repeatation of the batch procedure using 1.0$\times$10$^{-9}$ g g$^{-1}$ Ba solution. The recovery rate was stable in all batch procedures with variation of $<5$\%.

Next, 500 g of the Gd$_2$(SO$_4$)$_3{\cdot}$8H$_2$O powder was directly measured using the HPGe detector for the reference value. 
The measurement was performed for 9.3 days, and the concentration of $^{226}$Ra in the Gd$_2$(SO$_4$)$_3{\cdot}$8H$_2$O was estimated to be 2.8$^{+1.2}_{-0.9}$ mBq (5.6$^{+2.4}_{-1.8}$ mBq/kg).  
Finally, $^{226}$Ra in 100 g and 300 g of Gd$_2$(SO$_4$)$_3{\cdot}$8H$_2$O was extracted using the resin.
The concentration of the extracted $^{226}$Ra by the resin was evaluated using the HPGe detector. 
For the procedure blank, the batch procedure with 900 g of 10\% HNO$_3$ solution without Gd$_2$(SO$_4$)$_3{\cdot}$8H$_2$O was also performed. 

Table \ref{tab:results} shows the summary of the measurements. 
By taking the average of the values for a period of sixty minutes in Figure \ref{fig:Ba} and Table \ref{tab:results}, the averaged recovery rate of Ba was estimated to be 81$\pm$4(RMS)\%. 
The uncertainties of the estimated concentration for $^{226}$Ra mainly came from the calibration of the HPGe detector (+30\% or $-10$\%) and statistics. 
Comparing the 100 g, 300 g, and 500 g of samples with or without the chemical extraction, the converted concentration of Ra with unit of mBq/100 g (Gd$_2$(SO$_4$)$_3{\cdot}$8H$_2$O) in Table \ref{tab:results} are consistent with each other. 
This indicates that the concentration of Ra can be evaluated using the resin with correction of the recovery rate of Ba.

\begin{table}[htbp]
\begin{center}
\caption{The result of the HPGe measurement with or without the resin. For the concentration of $^{226}$Ra for 100 and 300 g samples, the value of the blank is subtracted and correction of the recovery rate is applied.}
\tabcolsep7pt\begin{tabular}{c|cccc}\hline
\multirow{2}{*}{Sample}  & \multirow{2}{*}{Procedure blank}  & 100 g  & 300 g & 500 g \\
 & & with the resin & with the resin & without the resin \\ \hline
Recovery rate & \multirow{2}{*}{82$\pm$2} & \multirow{2}{*}{86$\pm$1} & \multirow{2}{*}{81$\pm$2} & \multirow{2}{*}{-} \\
of Ba (\%) & & & & \\ \hdashline
Measurement & \multirow{2}{*}{10.0} & \multirow{2}{*}{5.8} & \multirow{2}{*}{9.6} & \multirow{2}{*}{9.3} \\
period (days) & & & & \\ \hdashline
Concentration  & \multirow{2}{*}{0.3$\pm$0.2} & \multirow{2}{*}{0.9$\pm$0.5} & \multirow{2}{*}{1.9$^{+0.7}_{-0.4}$} & \multirow{2}{*}{2.8$^{+1.2}_{-0.9}$} \\
 of $^{226}$Ra (mBq) & & & & \\ \hdashline
Converted concentration & \multirow{2}{*}{-} &\multirow{2}{*}{0.9$\pm$0.5} & \multirow{2}{*}{0.6$^{+0.2}_{+0.1}$} & 0.6$\pm$0.2 \\
of $^{226}$Ra (mBq/100 g) & & & & (reference) \\ \hline
\end{tabular}
\label{tab:results}
\end{center}
\end{table}

\section{Conclusion and foreseen improvement}

The new method of the chemical extraction for $^{226}$Ra from Gd$_2$(SO$_4$)$_3{\cdot}$8H$_2$O to minimize the problems such as the limit of the sample space and the detection efficiency coming from the geometry of the HPGe detector was developed using the molecular recognition resin ``AnaLig-Ra01". 
Using the developed batch method, $^{226}$Ra could be extracted from Gd$_2$(SO$_4$)$_3{\cdot}$8H$_2$O and concentrated to the resin with a recovery rate of 81$\pm$4\%. 
The result of the measurement with or without the batch procedure were consistent with each other within the uncertainties. 

For SK-Gd, the maximum allowed level of $^{226}$Ra in Gd$_2$(SO$_4$)$_3{\cdot}$8H$_2$O is 0.5 mBq/kg, which is an order lower than the sample used in this study. 
Therefore, more than 1 kg of the purer Gd$_2$(SO$_4$)$_3{\cdot}$8H$_2$O should be concentrated by repeat of the procedure to maximize the concentration rate and the sensitivity of the HPGe. 
For example, if the developed batch method was performed to 1 kg of Gd$_2$(SO$_4$)$_3{\cdot}$8H$_2$O with the concentration of 0.5 mBq (Ra)/kg, this situation would correspond to the sample with 100 g of hot spring water doped Gd$_2$(SO$_4$)$_3{\cdot}$8H$_2$O as shown in Table \ref{tab:results}. 


%


\ack
This work was supported by the JSPS KAKENHI Grants Grant-in-Aid for Scientific Research on Innovative Areas No. 26104004 and 26104006, Grant-in-Aid for Specially Promoted Research No. 26000003, Grant-in-Aid for Young Scientists No. 17K14290, and Grant-in-Aid for JSPS Research Fellow No. 18J00049. 
We thank the company who provided the Gd$_2$(SO$_4$)$_{3}{\cdot}$8H$_2$O sample. 



\end{document}